\documentclass[a4paper,12pt]{article}
\usepackage{graphicx}
\usepackage{epstopdf}
\usepackage{amsfonts,amsmath,amssymb}
\usepackage{hyperref}

\usepackage{epsfig,multicol}

\newcommand\fverb{\setbox\pippobox=\hbox\bgroup\verb}
\newcommand\fverbit{\egroup\item[\fbox{\unhbox\pippobox}]}

\newbox\pippobox

\begin{document}
\title{\bf Trace anomaly and invariance under transformation of units }
\author{ Nadereh Namavarian\thanks{Electronic address: n\_namavarian@sbu.ac.ir}\,\,
\\
{\small Department of Physics, Shahid Beheshti University, G.C.,
        Evin, Tehran 19839, Iran} }

 \maketitle
\begin{abstract}
\noindent
 Paying attention to conformal invariance as the invariance under 
local transformations of  units of measure, we take 
a conformal invariant quantum field as a quantum matter theory in which 
one has the freedom to choose the values of units of mass, length 
and time arbitrarily at each point. To be able to have this view,
it is necessary that the background 
on which the quantum field is based be conformal invariant as well. Consequently, 
defining the unambiguous expectation value of the energy-momentum tensor 
of such a quantum field through the Wald renormalizing prescription 
necessitates breaking down the conformal symmetry of the background. 
Then, noticing the field equations suitable for describing the back-reaction 
effect, we show that the 
existence of ``trace anomaly", known for indicating the 
brokenness of conformal symmetry in quantum field theory, 
can also indicate the above ``gravitational" 
conformal symmetry brokenness.

\end{abstract}
%

\vspace*{25pt}
\section {Introduction}
\indent

It is known that the renormalization of energy-momentum tensor in curved spacetimes has some ambiguities. In this context, different renormalizing methods have been developed, such as point separation~\cite{Chris, Adler}, dimensional regularization~\cite{Deser,Brown} and zeta-function regularization~\cite{Dowker,Hawking}, however, the results do not agree with each other completely. To provide a means for determining which renormalization prescription is correct, an axiomatic approach  was developed by Wald~\cite{Wald 1}. 
 In this approach, one tries to derive the expression of the renormalized energy-momentum tensor from the five axioms that it must satisfy, and it is shown that there can be at most one expression for $\langle {T_{\mu \nu }}\rangle$ which is compatible with these axioms (the uniqueness theorem~\cite{Wald 1}). However, the prescription for computing energy-momentum tensor proposed by Wald can satisfy only four of these axioms, and hence, he could not succeed in demonstrating the existence of a renormalized energy-momentum tensor compatible with all of the axioms~\cite{Wald 1}. Nevertheless, this axiomatic view helped to realize that all the employed renormalization schemes are correct in physical terms and the differences between them can be understood on the conceptual grounds. 

An interesting result in many of the works on renormalizing energy-momentum tensor is that, in the case of conformal invariant quantum fields, there exist non-zero values for the trace (in contrary to the classical conformal invariant fields where the trace of the energy-momentum tensor is identically zero). In addition, the uniqueness theorem of~\cite{Wald 1} establishes that no prescription consistent with the axioms 1-4 will yield traceless energy-momentum tensors for conformal invariant fields~\cite{Wald 2}, and hence, the existence of trace anomaly is inevitable in quantum field theory. Therefore, due to this fact, i.e., inevitable presence of trace anomaly and violation of the vanishing trace condition ($T^{\mu}_{\mu}=0$) on classical conformal invariant fields, it is well-known that the conformal symmetry is broken at quantum level.  

In this work, we consider the conformal symmetry of a quantum field as the freedom to choose the values of the units of length, time and mass arbitrarily at each point\footnote{Throughout this work, we assume $\hbar  = 1 = c$}. 
Then we show that, having this view toward the conformal symmetry of a field, and applying the Wald renormalization prescription to compute the energy-momentum tensor, the non-vanishing trace can also indicate a ``gravitational" conformal symmetry brokenness.
In the next section, a brief review is given on conformal invariance. In section 3, we introduce the field equations suitable for describing the back-reaction effect of the above mentioned conformal invariant quantum field, and investigate what the existence of trace anomaly can imply in this view.

\section {Conformal Invariance}
\indent

Conformal invariance\footnote{This phrase is usually used also for invariance under the action of the conformal group C(1,3). However, by conformal invariance we only mean the invariance under Weyl rescaling.} was first introduced into physics in 1919 through the Weyl geometry ~\cite{Weyl}. In an attempt to unify gravity and electromagnetic interaction, Weyl proposed a theory based on the idea that the unit of length is something arbitrary and can vary from point to point. Regarding the fact that conformal transformations of metric can be interpreted as local changes of the unit of length, Weyl realized the idea by generalizing Riemannian geometry to a geometry with conformal invariant connections through introducing an additional compensating vector field.  He tried to interpret the introduced field as the electromagnetic potential. However, this interpretation was not successful and it turned out that this field interacts in the same manner with both particles and antiparticles. Eventually, the Weyl theory was not accepted as a suitable theory for describing the electromagnetic interaction, and even, was rejected by Weyl himself~\cite{STM}. However, conformal invariance survived in physics as a principle which requires the physical laws to be invariant under local changes of the units of measure employed~\cite{Dicke}, and is of ever-increasing attention in modern physics.
 
It is evident that, to be able to have the above view toward the conformal invariance of a matter field, the gravitational background is needed to be conformal invariant in order that the dynamic of metric can permit local transformations of units of measure (i.e., conformal transformations of metric).
However, as we shall see in the next section, in the case of a ``quantum" matter field, the conformal symmetry of the background cannot be exact. Because one needs to break the symmetry in order to define the expectation value of the energy-momentum tensor.

\section {Back-reaction effect and arbitrariness in the values of units}
\indent

Classically, it is assumed that, matter influences gravity through its energy-momentum tensor. Hence, it is natural to expect a quantum field to have a back-reaction effect on the spacetime geometry as well. Under appropriate circumstances (particularly, when the fluctuations in $T_{\mu \nu }$ are sufficiently small compared with $T_{\mu \nu }$, and curvatures are small compared with Planck scale) the spacetime is still treated classically, and the back-reaction effect is described by the semi-classical equation

\begin{equation}\label{SCEE}
{G_{\mu \nu }} = 8\pi G \langle {T_{\mu \nu }}\rangle,
\end{equation}
 where the left hand side is given by the Einstein tensor ${G_{\mu \nu }} = {R_{\mu \nu }} - {\textstyle{1 \over 2}}R{g_{\mu \nu }}$, $G$ denotes the Newton gravitational constant, and the right hand side represents the expectation value of energy-momentum tensor in a given quantum state. In addition, many possibilities exists for deriving such a field equation via a suitable semi-classical limit from a quantum theory of gravity, e.g. see Ref~\cite{Flanagan}. Treating the back-reaction effect in this semi-classical approximation, the spacetime metrics and the quantum field states, which are dynamically possible, should be the ones satisfying the field equation~(\ref{SCEE}).

We take the right hand side of equation~(\ref{SCEE}) to be the expectation value of the energy-momentum tensor associated with a conformal invariant quantum field\footnote{We do not consider any special model for the quantum field. The results are the same for all conformal invariant scalar, spinor or vector fields.}. As mentioned above, to be able to take the conformal symmetry of the field as arbitrariness in the values of local units of measure, conformal invariance of the gravitational background is needed. However, regarding the left hand side of equation~(\ref{SCEE}), the background is described by general relativity which is not invariant under conformal transformations. To remedy this issue, one can generalize the Einstein-Hilbert action to 
\begin{equation}\label{action}
{S_[\psi ,g]} = \frac{1}{2}\int {{d^4}x\sqrt { - g} \left( {\frac{1}{6}R{\psi ^2} + {\partial _\mu }\psi {\partial ^\mu }\psi } \right)},
\end{equation}
in which $\psi$ is a scalar field that has been substituted for the gravitational constant (as the gravitational constant is a ``dimensional" constant, demanding the values of units to be arbitrary at each point results in replacing it with a scalar field). In addition, the coefficients have been chosen in a way that the action be invariant under the transformations  
\begin{equation}
{g_{\mu \nu }} \to {\Omega ^2}(x){g_{\mu \nu }}
\qquad\qquad {\rm and} \qquad\qquad
\psi (x) \to {\Omega ^{ - 1}}(x)\psi (x).
\end{equation}
Indeed, action~(\ref{action}) presents a scalar tensor theory in which gravity is described with both a metric tensor and a scalar field, like the Brans-Dicke theory~\cite{B-D}\footnote{This action might be considered the same as the Brans-Dicke action in the special case of $\omega  =  - \frac{3}{2}$. However, it should be noted that the scalar field in the Brans-Dicke theory has a different meaning than the one in~(\ref{action}). In this theory, the value of the gravitational constant varies with position as the consequence of Mach's principle, but, in~(\ref{action}), it varies as the consequence of variation of units of measure from point to point.}. Variation of this action with respect to  $\psi$ and $g_{\mu \nu }$ gives, respectively
\begin{equation}\label{phi eq}
\Box\,\psi - \frac{1}{6}R\psi = 0 
\end{equation}
and

\begin{equation}\label{g eq}
\frac{1}{6}{\psi ^2}{G_{\mu \nu }} - \tau _{\mu \nu }^{[\psi ]} = 0, 
\end{equation}
where

\begin{equation}\label{EM of phi}
\tau^{[\psi]}_{\mu \nu}\equiv  - {\partial_\mu }\psi
{\partial_\nu }\psi + \frac{1}{2}{g_{\mu \nu }}{\partial_\rho
}\psi {\partial^\rho }\psi  - \frac{1}{6}\left({g_{\mu \nu
}}\Box\, - {\nabla_\mu }{\nabla_\nu }\right){\psi^2}.
\end{equation}
Therefore, taking action~(\ref{action}) as the dynamical theory of the background\footnote{More generally, one can also add a potential term of the form $\lambda {\psi ^4}$ to action~(\ref{action}), where $\lambda$ is a dimensionless constant. However, it does not make any change in the results.}, the field equation~(\ref{SCEE}) will be replaced by equation~(\ref{phi eq}) and 
\begin{equation}\label{g eq2}
\frac{1}{6}{\psi ^2}{G_{\mu \nu }} - \tau _{\mu \nu }^{[\psi ]} = \langle {T_{\mu \nu }}\rangle,
\end{equation}
where, as mentioned above, $\langle {T_{\mu \nu }}\rangle$ is the expectation value of the energy-momentum tensor of a conformal invariant quantum field\footnote{However, there is a problem with equations~(\ref{phi eq}) and~(\ref{g eq2}) that will be explained in the following.}. Now, we take the energy-momentum tensor computed through the Wald renormalizing prescription and investigate how imposing the axiom 5\footnote{``The energy-momentum expression should not contain local curvature terms depending on derivatives of the metric higher than second order."} on $\langle {T_{\mu \nu }}\rangle$ affects the field equations~(\ref{phi eq}) and~(\ref{g eq2}).

In the Wald prescription~\cite{Wald 1}, the expectation value of the energy-momentum tensor of a quantum field is computed as follows: the Hadamard solution $S_0(x,x')$ defined by setting ${W_0}=0$\footnote{A Hadamard solution (i.e., a solution with the singularity structure near the point ${x'}$ of the form 
\begin{equation}\label{}
S(x,x') = \frac{U}{\sigma } + V\ln \sigma  + W,
\end{equation}
where ${\sigma }$ is half of the square of the geodesic distance between ${x}$ and ${x'}$, and ${U}$, ${V}$ and ${W}$ are smooth functions of ${x}$.)
can be explicitly constructed by expanding ${V}$ and ${W}$ in powers of ${\sigma }$,

\begin{equation}
V(x,x') = \sum\limits_{l = 0}^\infty  {{V_l}(x,x'){\sigma ^l}}, 
\qquad,\qquad
W(x,x') = \sum\limits_{l = 0}^\infty  {{W_l}(x,x'){\sigma ^l}},
\end{equation}
and solving the equation for ${U}$ and recursion relation for ${V_l}$ and ${W_l}$ for a given ${W_0}$~\cite{Hadamard}.} should be subtracted from the expectation value of the anticommutator of the field operator $G(x,x')$. Then, the resulting expression $F(x,x') = G(x,x') - S_0(x,x')$ should be substituted for $G(x,x')$ in the formal point separation expression of ${T_{\mu \nu }}$.
Assuming that $G(x,x')$ has the Hadamard form, the coincidence limit $x' \to x$ exists and yields a ``finite" expression for ${T_{\mu \nu }}$. However, as this expression is not conserved (i. e., ${\nabla ^\mu }\langle {T_{\mu \nu }}\rangle = {\nabla _\nu }Q$, where $Q$ is a local curvature term), the prescription is modified by subtracting ``by hand" the local ``correction term", $Q{g_{\mu \nu }}$ from $\langle {T_{\mu \nu }}\rangle $. Thereby, this modified prescription yields a finite and conserved expression for $\langle {T_{\mu \nu }}\rangle $ in which the presence of the ``correction term" gives rise to the appearance of trace anomalies for the conformal invariant fields. However, there is an ambiguity due to the assumption that $G(x,x')$ is of the Hadamard form~\cite{Wald 2}. In the prescription above, the length scale, by which $ \sigma$ is measured, is arbitrary, and changing it by a factor $\lambda$ (i.e., $ {g_{\mu \nu }} \to {\lambda ^2}{g_{\mu \nu }}$), where $\lambda$ is a dimensionless constant, results in the addition of an ambiguity term to $T_{\mu \nu }$, that is $ \ln \lambda $ times a conserved local curvature term quadratic in curvature. In the conformally invariant case, this ambiguity term is of the form $({I_{\mu \nu }} - 3{J_{\mu \nu }})$, where

\begin{equation}\label{}
{I_{\mu \nu }} =  - 2{g_{\mu \nu }}{\nabla _\alpha }{\nabla ^\alpha }R - 2R{R_{\mu \nu }} + 2{\nabla _\mu }{\nabla _\nu }R + {\textstyle{1 \over 2}}{R^2}{g_{\mu \nu }},
\end{equation}
and
\begin{equation}\label{}
{J_{\mu \nu }} = {\textstyle{1 \over 2}}{g_{\mu \nu }}({R_{\alpha \beta }}{R^{\alpha \beta }} - {\nabla _\alpha }{\nabla ^\alpha }R) + {\nabla _\mu }{\nabla _\nu }R - {\nabla _\alpha }{\nabla ^\alpha }{R_{\mu \nu }} - 2{R^{\alpha \beta }}{R_{\alpha \mu \beta \nu }}.
\end{equation}
On the other hand, from the Wald axiomatic view, the correct expression for $\langle {T_{\mu \nu }}\rangle $ is the ``unique" one which satisfies his five axioms. However, it can be checked that the prescription above can only satisfy the axioms 1-4~\cite{Wald 1, Wald 3}. As any expression compatible with axioms 1-4 is unique up to a local curvature term~\cite{Wald 1}, hence, there is a renormalization freedom in the expression obtained by the above prescription. Taking dimensional consideration into account, it can be shown~\cite{Wald 2} that this renormalization freedom term is of the forms $I_{\mu \nu }$ and $J_{\mu \nu }$ for the massless fields. 
Thus, to impose the axiom 5 and obtain the unique correct expression for energy-momentum tensor, that we denote by ${\langle {T_{\mu \nu }}\rangle _U}$, one needs to introduce a length unit into the theory~\cite{Wald 2}. On the other hand, specifying a length unit on a conformal invariant background (i.e., a background on which there exists the freedom to choose the units of measure arbitrarily) amounts to breaking down the conformal symmetry. Therefore, in the case of the above mentioned conformal invariant quantum field, imposing the axiom 5 on the expression of $\langle {T_{\mu \nu }}\rangle $  results in breaking down the conformal symmetry of the background. 

To break the conformal symmetry of the background, one can introduce the term $m^{2}{\psi^{2}}$ into action~(\ref{action}):

\begin{equation}\label{broken-action}
{S_[\psi , g] } = \frac{1}{2}\int {{d^4}x\sqrt { - g} \left(
\frac{1}{6}R{\psi^2} +{\partial_\mu }\psi {\partial^\mu
}\psi+m^{2}{\psi^{2}}  \right)},
\end{equation}
where $m$ is a constant with the dimension of mass. Consequently, the field equations~(\ref{phi eq}) and~(\ref{g eq}) turn into 

\begin{equation}\label{broken-phi}
\Box\,\psi - \frac{1}{6}R\psi - m^{2}\psi = 0
\end{equation}
and

\begin{equation}\label{broken-g}
\frac{1}{6}{\psi ^2}{G_{\mu \nu }} - \frac{1}{2}{m^2}{\psi ^2}{g_{\mu \nu }}-\tau_{\mu \nu }^{[\psi ]}  =0.
\end{equation}
To take action~(\ref{broken-action}) as the background on which the quantum field back-reacts through ${\langle {T_{\mu \nu }}\rangle _U}$, one can identify $m$ with the inverse of the length unit chosen to obtain ${\langle {T_{\mu \nu }}\rangle _U}$\footnote{In practice, there is no preferred length unit associated with a massless field or with the axioms~\cite{Wald 2, Wald 3}. However, in the above case that the background is conformal invariant, one can propose a length unit relevant to the scale chosen for the background (see Ref.\ \cite{N-F} for a discussion on this subject).}, that we denote by $m_0$. Also, introducing a length unit reduces the scalar field $\psi$ (the varying gravitational constant) to a constant value, that we denote by ${\psi _0}$, and thus, the field equations associated with the back-reaction effect will be the constraint $R=-6{m_0}^2$ and the equation
 
\begin{equation}\label{broken-g 2}
{G_{\mu \nu }} - 3{{m_0}^2}{g_{\mu \nu }} = 6{\psi _0}^{ - 2}{\langle {T_{\mu \nu }}\rangle _U}. 
\end{equation}

Now, it is worth noticing the point that equations~(\ref{broken-phi}) and~(\ref{broken-g}) are not independent of each other. Taking the trace of~(\ref{broken-g}) yields
\begin{equation}
\psi (\Box\,\psi - \frac{1}{6}R\psi - 2m^{2}\psi) =0,
\end{equation}
which (according to equation~(\ref{broken-phi})) implies that, in the absence of a matter field whose energy-momentum tensor is not traceless, the field equations obtained from action~(\ref{broken-action}) cannot be consistent with each other unless one sets $m=0$. In other words, to be able to break the conformal symmetry (i.e., to introduce the non-zero dimensional constant $m$ into action~(\ref{action})), it is necessary that the trace of the energy-momentum tensor of the conformal invariant quantum field be non-zero, i.e., the trace anomaly exists. 
On the other hand, taking the trace of~(\ref{g eq2}) yields

\begin{equation}\label{zero trace}
\psi( \Box\,\psi - \frac{1}{6}R\psi )= \langle {T^{\mu}_{\mu}}\rangle,
\end{equation}
which (according to equation~(\ref{phi eq})) shows that the consistency relation $\langle {T^{\mu}_{\mu}}\rangle=0$ is a constraint that is imposed by the conformal symmetry on the behavior of the matter system that can be coupled to this background, and hence, the presence of the non-vanishing trace is in contradiction with the conformal symmetry of the background.
Therefore, one can say that the existence of trace anomaly is a necessary and sufficient condition for the conformal symmetry of the background to be broken. In other words, the existence of trace anomaly can be considered as the indication of conformal symmetry brokenness of the gravitational background. 
In addition, as all the terms in trace anomaly are purely geometrical and independent of the quantum state\footnote{Trace anomaly is generally of the form $\alpha {C^{\mu \nu \rho \sigma }}{C_{\mu \nu \rho \sigma }} + \beta ({R^{\mu \nu }}{R_{\mu \nu }} - \frac{1}{3}{R^2}) + \gamma {\nabla _\mu }{\nabla ^\mu }R$, where $C^{\mu \nu \rho \sigma }$ is the Weyl tensor, and $\alpha$, $\beta$ and $\gamma$ are constants which depend on the field under consideration}, the relation ${\langle {T^{\mu}_{\mu}}\rangle _U} = - {m_0}^{2}{\psi _0}^{ 2}$ (obtained from taking the trace of~(\ref{broken-g 2}) and noticing the constraint $R=-6{m_0}^2$), implies that the presence of trace anomaly leads to the imposition of a constraint on the geometry of background spacetime.

\section {Conclusion}
\indent

Considering conformal invariance as the invariance under local transformations of units of measure, we have introduced the field equations suitable for describing the back-reaction effect of a conformal invariant quantum field. Then, taking the expectation value of energy-momentum tensor computed through the Wald renormalization prescription, we have investigated how the imposition of the 5th Wald axiom on the expression of $\langle {T_{\mu \nu }}\rangle$ affects the field equations of the back-reaction. Finally, we have deduced from the equations that, although the dynamic of the quantum field is invariant under the transformation of units, the existence of trace anomaly implicitly indicates that the preference of a specific unit of length is possible (i.e., conformal symmetry of the background is broken). In addition, preferring a length unit causes the trace anomaly to appear as a constraint on the geometry of the background spacetime. 

\section*{Acknowledgements}
\indent
I wish to thank Professor Mehrdad Farhoudi for his hospitality and useful comments. I also wish to thank Thomas-Paul Hack for helpful discussions, and his comments on this work. I thank the Research Office of Shahid Beheshti University for research facilities.


\end{document}